\documentclass{article}
\usepackage[T1]{fontenc} %
\usepackage[utf8]{inputenc} %
\usepackage{ismir,amsmath,amssymb,cite,url}
\usepackage{graphicx}
\usepackage{color}
\usepackage{textcomp}
\usepackage{booktabs}
\usepackage{multirow}
\usepackage{etoolbox,siunitx}
\usepackage{enumitem}

\renewcommand{\bfseries}{\fontseries{b}\selectfont}
\newrobustcmd{\B}{\bfseries}

\DeclareMathOperator*{\E}{\mathbb{E}}
\newcommand{\mb}{\mathbf}

\title{Multi-instrument Music Synthesis\\with Spectrogram Diffusion}

\multauthor
{Curtis Hawthorne \hspace{1cm} Ian Simon \hspace{1cm} Adam Roberts \hspace{1cm} Neil Zeghidour} { \bfseries{Josh Gardner$^{\star}$ \hspace{1cm} Ethan Manilow$^{\star}$\thanks{$^{\star}$ Work done as a Google Brain Student Researcher.} \hspace{1cm} Jesse Engel}\\
Google Research, Brain Team\\
{\tt\small \{fjord,iansimon,adarob,neilz,jesseengel\}@google.com} \\
\tt \small jpgard@cs.washington.edu, ethanm@u.northwestern.edu
}

\def\authorname{C.~Hawthorne, I.~Simon, A.~Roberts, N.~Zeghidour, J.~Gardner, E.~Manilow, and J.~Engel}

\usepackage[bookmarks=false,pdfauthor={\authorname},pdfsubject={\papersubject},hidelinks]{hyperref}
\usepackage[capitalise,noabbrev]{cleveref}

\begin{document}

\maketitle
\begin{abstract}
An ideal music synthesizer should be both interactive and expressive, generating high-fidelity audio in realtime for arbitrary combinations of instruments and notes. Recent neural synthesizers have exhibited a tradeoff between domain-specific models that offer detailed control of only specific instruments, or raw waveform models that can train on any music but with minimal control and slow generation. In this work, we focus on a middle ground of neural synthesizers that can generate audio from MIDI sequences with arbitrary combinations of instruments in realtime. This enables training on a wide range of transcription datasets with a single model, which in turn offers note-level control of composition and instrumentation across a wide range of instruments.  We use a simple two-stage process: MIDI to spectrograms with an encoder-decoder Transformer, then spectrograms to audio with a generative adversarial network (GAN) spectrogram inverter. We compare training the decoder as an autoregressive model and as a Denoising Diffusion Probabilistic Model (DDPM) and find that the DDPM approach is superior both qualitatively and as measured by audio reconstruction and Fréchet distance metrics. Given the interactivity and generality of this approach, we find this to be a promising first step towards interactive and expressive neural synthesis for arbitrary combinations of instruments and notes.
\end{abstract}

\begin{figure*}
 \centerline{
 \includegraphics[width=0.80\textwidth]{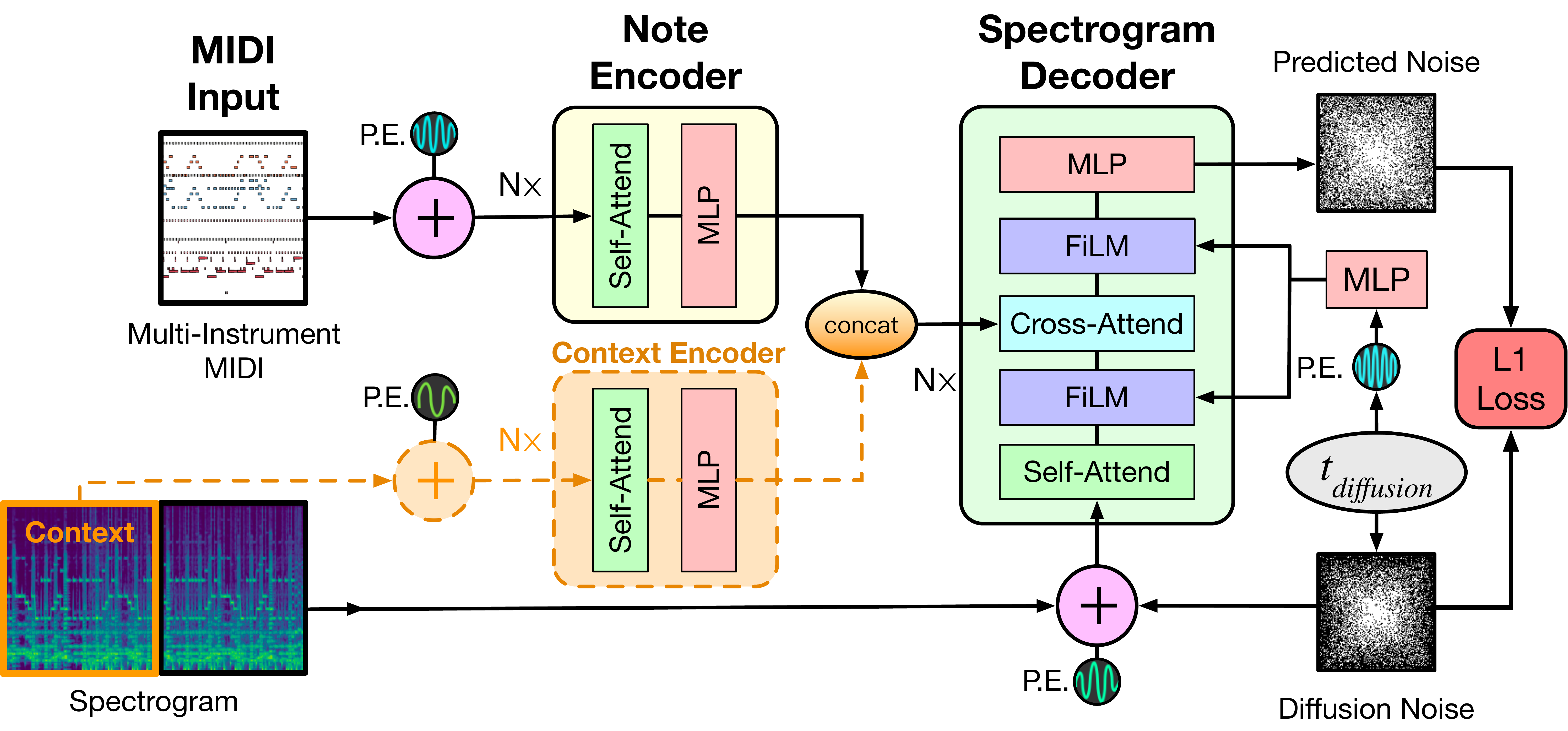}}
 \caption{Training configuration for our spectrogram diffusion model. The architecture is an encoder-decoder Transformer that takes a sequence of note events as input and outputs a spectrogram. We train the decoder stack as a Denoising Diffusion Probabilistic Model (DDPM)~\cite{ho2020denoising}, where the model learns to iteratively refine Gaussian noise into a target spectrogram (\cref{fig:diffusion}).
 We generate $\sim$5 second spectrogram segments, and to ensure a smooth transition between these segments we (optionally) encode the previously generated segment in a second encoder stack.
 At inference time, a generated spectrogram is inverted to a waveform using a model similar to MelGAN~\cite{kumar2019melgan}.
 ``P.E.'' means Positional Encoding.}
 \label{fig:architecture}
\end{figure*}

\section{Introduction}\label{sec:introduction}

Neural audio synthesis of music is a uniquely difficult problem due to the wide variety of instruments, playing styles, and acoustic environments. 
Among current approaches, there is often a trade-off between interactivity and generality.
Interactive models, such as DDSP \cite{Engel2020DDSP:,wu2021midi}, offer realtime synthesis and fine-grained control, but only for specific types of instruments with domain-specific training information. On the other hand, models like Jukebox \cite{dhariwal2020jukebox} are much more general and allow for training on any type of music, but are several orders of magnitude slower than realtime and offer more limited and global control, such as lyrics and global style (though with some early MIDI conditioning experiments).

Natural language generation and computer vision have seen dramatic recent progress through scaling generic encoder-decoder Transformer architectures~\cite{roberts2022scaling, flamingo}. MT3~\cite{gardner2022mt, hawthorne2021sequence} demonstrated that such an approach can be adapted to Automatic Music Transcription, transforming spectrograms to variable-length sequences of notes from arbitrary combinations of instruments. 
This generic approach enables training a single model on a wide variety of datasets: anything with paired audio and MIDI examples. 

In this paper, we explore the inverse problem: finding a simple and general method to transform variable-length sequences of notes from arbitrary combinations of instruments into spectrograms.
By pairing this model with a spectrogram inverter, we find that we are able to train on a wide variety of datasets and synthesize audio with note-level control over both composition and instrumentation.

To summarize, our contributions include:
\begin{itemize}
    \item Demonstrating that the generic encoder-decoder Transformer approach used for multi-instrument transcription can likewise be adapted for multi-instrument audio synthesis.
    \item Realtime synthesis enabling interactive note-level control over both composition and instrumentation, by pairing a MIDI-to-spectrogram Transformer model with a GAN spectrogram inverter, and training on a wide range of paired audio/MIDI datasets (synthetic and real) with diverse instruments.
    \item Adapting DDPM decoding for arbitrary length synthesis without edge artifacts, through additional autoregressive conditioning on segments ($\sim$5 seconds) of previously generated audio.
    \item Quantitative metrics and qualitative examples demonstrating the advantages of segment-wise diffusion decoders over frame-wise autoregressive decoders for continuous spectrograms.
    \item Source code and pretrained models\footnote{\url{https://github.com/magenta/music-spectrogram-diffusion}}.
\end{itemize}

\section{Related Work}

Neural audio synthesis first proved feasible with autoregressive models of raw waveforms such as WaveNet and SampleRNN~\cite{vanwavenet, mehri2016samplernn}. These models were adapted to handle conditioning from latent variables~\cite{engel2017neural, universalmusic} or MIDI notes~\cite{hawthorne2018enabling, manzelli2018conditioning, kim2019neural, cooper2021text}, but suffer from slow generation due to needing to run a forward pass for every sample of the waveform. Real-time synthesis often requires the use of specialized CPU and GPU kernels~\cite{kalchbrenner2018efficient, rtnsynth}.
These models also have limited temporal context due to the high sample rate of audio (e.g., 16kHz or 48kHz), where even the largest receptive fields (thousands of timesteps) can equate to less than a second of audio.

Approaches to overcome these speed limitations of waveform autoregression have focused on generating audio directly with a single forward pass. Architectures commonly either use GANs~\cite{donahue2018adversarial, engel2018gansynth, greshler2021catchawaveform, morrison2022chunked}, controllable differentiable DSP elements such as oscillators and filters~\cite{Engel2020DDSP:, wu2021midi, hpn, ddsp-inv, castellontowards, neuralsourcefilter}, or both~\cite{caillon2021rave, caillon2022streamable}. 
These models often focus on limited domains such as generating a single instrument/note/voice at a time~\cite{defossez2018sing, manzelli2018conditioning, kim2019neural, wu2021midi}.
Here, we focus our search on architectures capable of both realtime synthesis, note-level control, and synthesizing mixtures of multiple polyphonic instruments at once.

Researchers have also overcome temporal context limitations of waveform autoregression by adopting a multi-stage approach, first creating coarser audio representations at a lower sample rate, and then modeling those representations with a predictive model before decoding back into audio. 
For example, Jukebox and Soundstream~\cite{dhariwal2020jukebox, zeghidour2021soundstream, hawthorne2022general} use a Transformer to autoregresively model the discrete vector-quantized codes~\cite{vqvae} of a base waveform autoencoder.

Tacotron architectures~\cite{wang2017tacotron, tacotron2,weiss2021wave} have demonstrated that straightforward spectrograms can be an effective audio representation for multi-stage generation, first autoregressively generating continuous-valued spectrograms, and then synthesizing waveforms with a neural vocoder. 
The success of this approach led to a flurry of research into spectrogram inversion models, including streamlined waveform autoregression, GANs, normalizing flows, and Denoising Diffusion Probabilistic Models (DDPMs)~\cite{kalchbrenner2018efficient, kumar2019melgan, prenger2019waveglow, ho2020denoising, kong2020diffwave, chen2020wavegrad}. 
Our approach here is inspired by Tacotron. We use an autoregressive spectrogram generator paired with a GAN spectrogram inverter as a baseline, and further improve upon it with a DDPM spectrogram generator.

DDPMs and Score-based Generative Models (SGMs) have proven well-suited to generating audio representations and raw waveforms. Speech researchers have demonstrated high-fidelity spectrogram inversion~\cite{kong2020diffwave, chen2020wavegrad}, text-to-speech~\cite{popov2021grad, jeong2021diff}, upsampling~\cite{lee2021nu}, and voice conversion~\cite{kameoka2020voicegrad, popov2021diffusion}. Musical applications include singing synthesis of individual voices~\cite{diffsinger, liu2021diffsvc}, drums synthesis~\cite{rouard2021crash}, improving the quality of music recordings~\cite{kandpal2022music}, symbolic note generation~\cite{mittal2021symbolic}, and unconditional piano generation~\cite{goel2022s}. Similar to singing synthesis, here we investigate DDPMs for spectrogram synthesis, but focusing on arbitrary numbers and combinations of instruments.

\section{Architecture}

\begin{figure*}
 \centerline{
 \includegraphics[width=\textwidth]{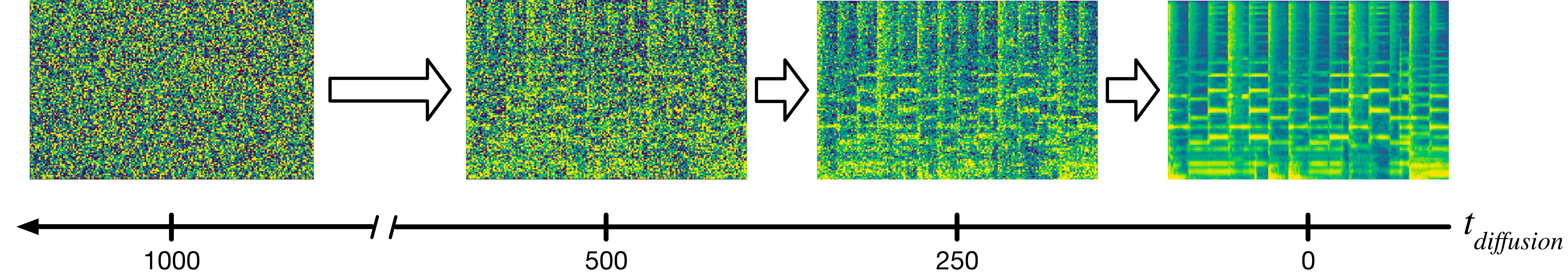}}
 \caption{Our decoder stack is trained as a Denoising Diffusion Probabilistic Model (DDPM)~\cite{ho2020denoising}. The model starts with Gaussian noise as input and is trained to iteratively refine that noise toward a target, conditioned on a sequence of note events and the spectrogram of the previously rendered segment. This figure illustrates the diffusion process for one example segment.}
 \label{fig:diffusion}
\end{figure*}

We approach the problem of audio synthesis using a two-stage system. The first stage consists of a model that produces a spectrogram given a sequence of MIDI-like note events representing any number of instruments. We then use a separate model to invert those spectrograms to audio samples. Our model for the first stage is an encoder-decoder Transformer architecture, shown in Figure~\ref{fig:architecture}. The encoder takes in a sequence of note events and, optionally, a second encoder uses an earlier part of the spectrogram as context. These embeddings are passed to a decoder, which generates a spectrogram corresponding to the input note sequence. Here, we explore training the decoder autoregressively (Section~\ref{sec:autoregressive}) or as a Denoising Diffusion Probabilistic Model (DDPM)~\cite{ho2020denoising} (Section~\ref{sec:diffusion}).

Our architecture uses the encoder-decoder Transformer from T5 \cite{raffel202Exploring} with T5.1.1\footnote{\url{https://github.com/google-research/text-to-text-transfer-transformer/blob/main/released_checkpoints.md\#t511}} improvements and hyperparameters. We use the same note event vocabulary and note sequence encoding procedure as MT3~\cite{gardner2022mt}. We find that training on a full song is prohibitive in terms of memory and compute due to the quadratic scaling of self-attention with sequence length, so we split note sequences into segments. Specifically, we train models with 2048 input positions for note events and 256 output positions for spectrogram frames. Each spectrogram frame represents 20 ms of audio, so each segment is $5.12$ seconds.

Rendering audio segments independently introduces the problem of how to ensure smooth transitions between the segments once they are eventually concatenated together to form a full musical piece. We address this problem by providing the model with the spectrogram from the previously rendered segment, meaning that this model is autoregressive at the segment level. This context segment has its own encoder stack with 256 input positions. As seen in \cref{fig:architecture}, the outputs of both encoder stacks are concatenated together as input for cross-attention in the decoder layers.

We use sinusoidal position encodings, as in the original Transformer paper \cite{vaswani2017attention}. However, we found better performance if we decorrelated the position encodings of each network, as they have different meanings. We decorrelate the encodings by applying a unique random channel permutation and phase offset for the sinusoids used in each of the encoder and decoder stacks.

\subsection{Autoregressive Decoder}
\label{sec:autoregressive}

As an initial approach, we took inspiration from Tacotron~\cite{wang2017tacotron, tacotron2}, and trained the decoder as an autoregressive model on the continuous spectrograms. In this setting, standard causal masking was applied throughout the self-attention layers. The inputs and outputs are continuous spectrogram frames, and the model was trained with a mean-squared error (MSE) loss on those frames. Mathematically, this is equivalent to training a continuous autoregressive model with a Gaussian output distribution with isotropic and fixed variance. We also tried training models using mixtures of several Gaussians, but they tended to be unstable during sampling.

For sampling we use a fixed variance ($0.2$ in units of log magnitude) that is added to the outputs of every inference step. In practice, this ``dithering'' reduced strong audio artifacts due to overly smooth outputs. 

Despite this, these baseline models still produce spectrogram outputs that tend to be blurry and contain jarring artifacts in some frames. We hypothesized an approach enabling incremental, bidirectional refinement, and model dependencies between frequency bins, could help resolve these issues.

\subsection{Diffusion Decoder}
\label{sec:diffusion}

Taking inspiration from recent successes in image generation such as DALL-E 2 and Imagen \cite{ramesh2022hierachical,saharia2022photorealistic}, we investigated training the decoder as a Denoising Diffusion Probabilistic Model (DDPM) \cite{ho2020denoising}.

DDPMs are models that generate data from noise by reversing a Gaussian diffusion process that turns an input signal $\mb{x}$ into noise $\mb{\epsilon} \sim \mathcal{N}(\mb{0}, \mb{I})$. This \emph{forward} process occurs over diffusion timesteps $t \in [0, 1]$ (not to be confused with time for the audio) resulting in noisy data $\mb{x_t} = \alpha_t \mb{x} + \sigma_t \mb{\epsilon}$, where $\alpha_t \in [0, 1]$ and $\sigma_t \in [0, 1]$ are noise schedules that blend between the original signal and the noise over the diffusion time.
In this work, the decoder, $\mb{\epsilon}_{\theta}$ with parameters $\theta$, is trained to predict the added noise given the noisy data by minimizing an L1 objective,
\begin{align}
    \E_{\mb{x}, \mb{c}, \mb{\epsilon}, t}  w_t || \mb{\epsilon}_{\theta}(\mb{x_t}, \mb{c}, t) - \mb{\epsilon} ||_1^1 
\label{eq:loss}
\end{align}
where $w_t$ is a loss weighting for different diffusion timesteps and $\mb{c}$ is additional conditioning information for the decoder. Schedules for $w_t$, $\alpha_t$ and $\mb{\epsilon}_t$ are hyperparameters chosen to selectively emphasize certain steps in the reverse diffusion process.

During sampling, we follow the \emph{reverse} diffusion process by starting from pure independent Gaussian noise for each frame and frequency bin and iteratively use noise estimates to generate new spectrograms with gradually decreasing noise content. Full details of this process can be found in Ho et al.~\cite{ho2020denoising} and source code is provided for clarity\footnote{\url{https://github.com/magenta/music-spectrogram-diffusion}}. 
A visualization of the forward and reverse diffusion process can be seen in \cref{fig:diffusion}.

Because the diffusion process expects inputs to be in the range $[-1, 1]$, we scale spectrograms used for training targets and audio context to that range before using them. During inference, we scale model outputs back to the expected spectrogram range before converting them to audio.

During training, we used a uniformly-sampled noise time step in $[0, 1]$ for each training example and run the diffusion process forward to that step with randomly sampled Gaussian noise. The noisy example is then used for input, and the model is trained to predict the Gaussian noise component with \cref{eq:loss}. During inference, we run the diffusion process in reverse using 1000 linearly spaced steps in $[0, 1]$.
    
Based on the implementation of Imagen \cite{saharia2022photorealistic}, we also incorporated several recent DDPM improvements, including using logSNR \cite{kingma2021variational,salimans2022progressive}, a cosine noise schedule \cite{nichol2021Improved} of $\textrm{cos}(t * \pi / 2)^2$, and Classifier-Free Guidance \cite{ho2021classifierfree} during sampling.
In particular, we found that using Classifier-Free Guidance during sampling led to samples that were less noisy and more crisp sounding. After a coarse sweep of values, we train with a conditioning dropout probability of $0.1$ and sample with a conditioned sample weight of $2.0$.

Many diffusion models use a UNet \cite{ronneberger2015u} architecture, but in order to stay close to the architecture used in MT3, we used a standard Transformer decoder with no causal masking. We incorporate the noise time information by first converting the continuous time value to a sinusoid embedding using the same method as the position encodings in the original Transformer paper \cite{vaswani2017attention} followed by an MLP block. That embedding is then incorporated at each decoder layer using a FiLM \cite{perez2018film} layer before the self-attention block and after the cross-attention block. The outputs of note events and spectrogram context encoder stacks are concatenated together and incorporated using a standard encoder-decoder cross-attend layer.

\subsection{Spectrograms to Audio}
\label{sec:spectrogram_inversion}

To translate the model's magnitude spectrogram output to audio, we use a convolutional spectrogram inversion network as proposed in MelGAN \cite{kumar2019melgan}. In particular, we base our implementation on the more recent SEANet \cite{tagliasacchi2020seanet} and SoundStream \cite{zeghidour2021soundstream}. This model first applies a 1D-convolution with kernel size 3 and stride 1. It then cascades four blocks of layers. Each block is composed of a transposed convolution followed by three residual units, applying a dilated convolution with kernel size 3 and dilation rate 1, 3, and 9 respectively. ELUs \cite{elu} are used after each convolution. As in \cite{zeghidour2021soundstream}, the model is trained with a mix of three loss functions. The first is a multi-scale spectral reconstruction loss, inspired by \cite{gritsenko2020spectral}. The second and third losses are an adversarial loss and a feature matching loss, computed with two discriminators, one STFT-based and one waveform-based. We train this spectrogram inverter on an internal dataset of 16k hours of uncurated music with Adam \cite{Kingma2015AdamAM} for 1M steps with a batch size of 128.

The spectrograms used for training both the spectrogram inverter and the synthesis models used audio with a sample rate of 16 KHz, an STFT hop size of 320 samples (20 ms), a frame size of 640, and 128 log magnitude mel bins ranging in frequency from 0 to 8 KHz.

\section{Datasets}

Our architecture is flexible enough to train on any dataset containing paired audio and MIDI examples, even when multiple instruments are mixed together in the same track (e.g., individual instrument stems are unavailable). Thus, we are able to use all the same datasets used to train the MT3 transcription model: MAESTROv3 \cite{hawthorne2018enabling} (piano), Slakh2100 \cite{manilow2019cutting} (synthetic multi-instrument), Cerberus4 \cite{manilow2020simultaneous} (synthetic multi-instrument), Guitarset \cite{xi2018guitarset} (guitar), MusicNet \cite{thickstun2017learning} (orchestral multi-instrument), and URMP \cite{li2019Creating} (orchestral multi-instrument).

We use the same preprocessing of these datasets as MT3, including the data augmentation strategy used for Slakh2100 where 10 random subsets of instruments from each track were selected to increase the number of tracks. We also use the same train/validation/test splits. Our coarse hyperparameter sweeps (e.g., for selecting Classifier-Free Guidance weighting) were done using only validation sets. Results are reported on the test splits, except for Guitarset and URMP which do not have a test split, and so the validation split was used for final results.

The datasets used in these models contain a wide variety of instruments, both synthetic and real. In order to simplify training, we map all MIDI instruments to the 34 Slakh2100 classes plus drums. These mappings cover all General MIDI classes other than ``Synth Effects'', ``Other'', ``Percussive'', and ``Sound Effects''. As a result, the trained model is capable of synthesizing arbitrary MIDI files while retaining clear instrument identity.

We use the same note event vocabulary as MT3, which is similar to MIDI events. Specifically, there are events for Instrument (128 values), Note (128 values), On/Off (2 values), Time (512 values), Drum (128 values), End Tie Section (1 value), and EOS (1 value). A full description of these events can be found in Section 3.2 of the MT3 paper \cite{gardner2022mt}. Because not all datasets include velocity information, we do not currently include velocity events (as in MT3) and rely on the model's decoder to produce natural sounding velocities given the music context.

Training on such a diverse set of examples gives the model flexibility during synthesis. For example, we can synthesize a track with realistic piano sounds learned from MAESTRO, orchestral instruments from URMP, and synthesizers and drum beats from Slakh.

\section{Experiments}
\label{sec:experiments}

\begin{table*}[t]
\sisetup{table-format=2.2,round-mode=places,round-precision=2,table-number-alignment = center,detect-weight=true,detect-inline-weight=math}
  \centering
  \begin{tabular}{l S S S S S S}
    \toprule
    \multicolumn{1}{c}{\multirow{2}{*}{Model}} & \multicolumn{2}{c}{VGGish~\cite{hershey17audiocnn}} & \multicolumn{2}{c}{TRILL~\cite{trill}} & {\multirow{2}{*}{{Transcription F1 $\uparrow$}}} & {\multirow{2}{*}{{RT Factor $\uparrow$}}} \\
    {} & {Recon. $\downarrow$} & {FAD $\downarrow$} & {Recon. $\downarrow$} & {FAD $\downarrow$} & {} & {} \\  
    \midrule
    Small Autoregressive & 3.70 & 1.03 & 7.27 & 0.39 & 0.31 & \B 10.55 \\
    Small w/o Context & 3.49 & 1.07 & 6.16 & 0.46 & 0.31 & 2.82 \\
    Small w/ Context & 3.56 & 1.10 & 6.40 & 0.53 & 0.25 & 3.07 \\
    Base w/ Context & \B 3.13 & \B 1.00 & \B 2.74 & \B 0.27 & \B 0.36 & 1.05 \\
    \midrule
    Ground Truth Encoded & 2.14 & 0.51 & 1.54 & 0.05 & 0.36 & 12.25 \\
    Ground Truth Raw & 0.00 & 0.00 & 0.00 & 0.00 & 0.63 & {--} \\
    \bottomrule
  \end{tabular}
  \caption{Synthesis model experiment results. Metrics are the mean across all evaluation datasets. ``Small'' and ``Base'' refer to model capacity (\cref{sec:experiments}). ``Ground Truth Encoded'' refers to the ground truth audio encoded by the spectrogram inverter (\cref{sec:spectrogram_inversion}) and represents an upper limit on synthesis model performance. ``Ground Truth Raw'' refers to the unprocessed ground truth audio and represents an upper limit on transcription model performance. Metrics are fully described in \cref{sec:metrics} and results are discussed in \cref{sec:results}.}
  \label{tab:results}
\end{table*}

\begin{figure}[t]
 \centerline{
 \includegraphics[width=0.95\columnwidth]{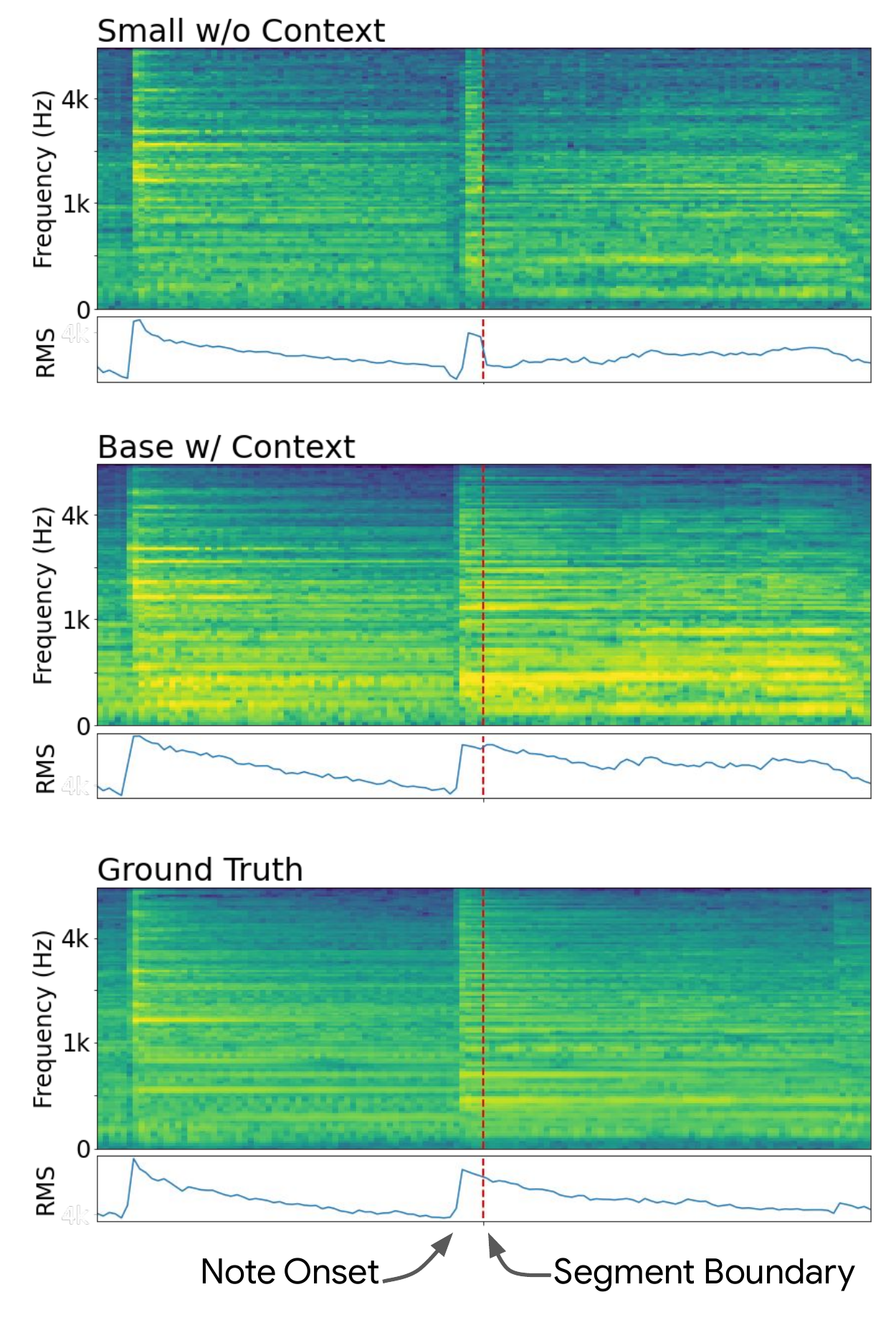}}
 \caption{The ``Small w/o Context'' model (top) does not receive the previously rendered segment as conditioning input and exhibits clear artifacts at segment boundaries (note the dip in RMS after the segment boundary). The ``Base w/ Context'' model (middle) has access to the previous segment as conditioning information and achieves smooth transition between segments. By repeating the process of rendering a segment and then feeding that segment in as context for the next segment, this model is capable of rendering arbitrary length pieces.}
 \label{fig:segment_boundary}
\end{figure}

We used the MT3 codebase as a starting point \cite{MT3Code}, and experiments were implemented using \texttt{t5x} for model training and \texttt{seqio} for data processing and evaluation \cite{roberts2022scaling}.

We used the T5.1.1 ``small'' or ``base'' Transformer hyperparameters. For ``small'', there are 8 layers on the encoder and decoder stacks, 6 attention heads with 64 dimensions each, 1024 MLP dimensions, and 512 embedding dimensions. For ``base'', there are 12 layers on the encoder and decoder stacks, 12 attention heads with 64 dimensions each, 2048 MLP dimensions, and 758 embedding dimensions. We used float32 activations for all models.

Using the datasets described above, we trained four versions of the synthesis model: an 45M parameter ``small'' autoregressive model with no spectrogram context, an 85M parameter ``small'' diffusion model with no context, a 104M parameter ``small'' diffusion model with context, and a 412M parameter ``base'' diffusion model with context. Training used a batch size of 1024 on 64 TPUv4s with a constant learning rate of $1\mathrm{e}^{-3}$ for 500k steps with the Adafactor \cite{shazeer2018Adafactor} optimizer. Depending on model size and hardware availability, training took 44--134 hours.

\subsection{Metrics}
\label{sec:metrics}

We evaluate model performance on the following metrics:

\begin{description}[leftmargin=0pt]
    \item[Reconstruction Embedding Distance] For this metric, we pass an individual audio clip and a synthesis model's reconstruction of that audio into a classification network and calculate the distance in the classification network's embedding between the two signals. Here, we report numbers from both VGGish~\cite{hershey17audiocnn} and TRILL~\cite{trill}.
    To compute the distance we use the Frobenius norm of the network's embedding layer, averaged over time frames.
    VGGish outputs 1 embedding per input audio second, and we use the model's output layer as the embedding layer. TRILL outputs $\sim$5.9 embeddings per input audio second, and we use the network's dedicated ``embedding'' layer. 
    
    \item[Fréchet Audio Distance (FAD)] FAD measures the perceptual similarity of the distribution of all of the model's output over the entire evaluation set of note events to the distribution of all of the ground truth audio \cite{kilgour2019frechet}.
    
    \item[MT3 Transcription] This metric measures how well the synthesis model is reproducing the notes and instruments specified in the input data. We pass the synthesis model output through a pretrained MT3 transcription model and compute an F1 score using the ``Full'' metric from the MT3 paper as calculated by \texttt{mir\_eval} \cite{raffel2014mir_eval}. A note is considered correct if has a matching note onset within $\pm 50$ ms, an offset within $0.2 \cdot \textrm{reference\_duration}$ or $50$ ms (whichever is greater), and has the exact instrument program number as the input data.
    
    \item[Realtime (RT) Factor] This measures how fast synthesis is compared to the duration of the audio being synthesized. For example, an RT Factor of 2 means that 2 seconds of audio can be created with 1 second of wall time compute. Inference is performed on a single TPUv4 with a batch size of 1. Here, we include both spectrogram inference and spectrogram-to-audio inversion.
\end{description}

The reconstruction and FAD metrics require embedding both model output and ground truth audio using a model sensitive to perceptual differences. Following the original FAD formulation, we use the VGGish model \cite{hershey17audiocnn}. To avoid any biases from that particular model, we also compute embeddings using the TRILL model \cite{trill}. Pretrained models for computing these embeddings were obtained from the VGGish\footnote{https://tfhub.dev/google/vggish/1} and TRILL\footnote{https://tfhub.dev/google/nonsemantic-speech-benchmark/trill/3} pages on TensorFlow Hub. Due to the large size of the datasets and the high compute and memory requirements for these metrics, we limit metrics calculation to the first 10 minutes of audio for each example.

\subsection{Results}
\label{sec:results}

We first evaluated a ``small'' autoregressive model (\cref{sec:autoregressive}). Initial results were promising, but quantitative and qualitative evaluation of results made it clear that this approach would not be sufficient.

We then investigated using a diffusion approach (\cref{sec:diffusion}). Results from a ``small'' model were impressive (nearly maxing out the Transcription metric), but we noticed abrupt timbre shifts and audible artifacts at segment boundaries, as illustrated in \cref{fig:segment_boundary}. This makes sense because the synthesis problem from raw MIDI is underspecified: input to the model specifies notes and instruments, but the model has been trained on a wide variety of sources, and the audio corresponding to a given note and instrument combination could just as likely be synthetic or real, played in a dry or reverberant room, played with or without vibrato, etc.; the model has to sample from this large space of possibilities independently for each segment.

To address this issue, we added a context spectrogram input encoder to encourage the model to be consistent over time. We first experimented with adding this context input to a ``small'' model, but got poor results, possibly due to insufficient decoder capacity for incorporating the additional encoder output. We then tried scaling up to a ``base'' model that included context. The additional model capacity resulted in generally higher audio quality and also did not have segment boundary problems. This is reflected in the metrics, where this model achieves the best scores by far. Also, even with the larger model size, the synthesis process is still slightly faster than realtime.

Qualitatively, we find that the model does an especially impressive job rendering instruments where the training data came from real audio recordings as opposed to synthetic instruments. For example, when given a note event sequence from the Guitarset validation split, the model not only accurately reproduces the notes but also adds fret and picking noises. Sequences rendered for orchestral instruments such as the ones found in URMP add breath noise and vibrato. The model also does a remarkably good job of rendering natural-sounding note velocities, even though no velocity information is present in our note vocabulary.

Results for these experiments are presented in \cref{tab:results}. Additional results, including audio examples demonstrating the ability to render out-of-domain MIDI files and modify their instrumentation, are in our online supplement\footnote{Online supplement and examples: \url{https://bit.ly/3wxSS4l}}. Results on a per-dataset basis are in \cref{sec:dataset_metrics}.

These results are encouraging, but there is still plenty of room for improvement. Particularly, the synthesized audio has occasional issues with inconsistent loudness and audio artifacts, and its overall fidelity does not match the training data. However, in all our experiments, we observed no overfitting on the validation sets, so we suspect that even larger models could be trained for higher fidelity audio.

Another limiting factor of our approach is the spectrogram inversion process, which represents an upper bound on audio quality. This is apparent in the Transcription F1 score, where our models have already achieved the maximum score possible, even though that score is only a little over half of what is achievable with raw audio.

\section{Conclusion}

This work represents a step toward interactive, expressive, and high fidelity neural audio synthesis for multiple instruments.
Our flexible architecture allows incorporating any training data with paired audio and MIDI examples. Improved automatic music transcription systems such as MT3~\cite{gardner2022mt} point toward being able to generate high quality MIDI annotations for arbitrary audio, which would greatly expand the available training data and range of instruments and acoustic settings.
Using generic Transformer encoder stacks also presents the possibility of utilizing conditioning information beyond note events and spectrogram context. For example, we could add finer grained conditioning such as control over note timbre, or more global controls such as free text descriptions.
This model is already slightly faster than realtime, but ongoing research into diffusion models shows promising directions for speeding up inference \cite{salimans2022progressive}, giving plenty of room for larger and more complicated models to remain interactive. This may be especially important as we explore options for higher quality audio output than our current spectrogram inversion process enables.

\section{Acknowledgements}

We would like to thank William Chan, Mohammad Norouzi, Chitwan Saharia, and Jonathan Ho for help with the diffusion implementation and Sander Dieleman for helpful discussions about diffusion approaches. Also thanks to Chin-Yun Yu for correcting a model parameter count.

\bibliography{ISMIRtemplate}

\clearpage
\newpage
\appendix
\onecolumn
\section{Dataset Metrics}
\label{sec:dataset_metrics}

The results presented in \cref{tab:results} are the mean of metrics across all datasets. Here we show the results on a per-dataset basis.

\begin{table*}[h]
\sisetup{table-format=2.2,round-mode=places,round-precision=2,table-number-alignment = center,detect-weight=true,detect-inline-weight=math}
  \centering
  \begin{tabular}{l S S S S S}
    \toprule
    \multicolumn{1}{c}{\multirow{2}{*}{Model}} & \multicolumn{2}{c}{VGGish~\cite{hershey17audiocnn}} & \multicolumn{2}{c}{TRILL~\cite{trill}} & {\multirow{2}{*}{{Transcription F1 $\uparrow$}}} \\
    {} & {Recon. $\downarrow$} & {FAD $\downarrow$} & {Recon. $\downarrow$} & {FAD $\downarrow$} & {} \\  
    \midrule
    Small Autoregressive & 2.93 & 1.01 & 2.91 & 0.27 & 0.47 \\
    Small w/o Context & 2.91 & 1.12 & 3.05 & 0.41 & 0.40 \\
    Small w/ Context & 3.02 & 1.16 & 3.44 & 0.52 & 0.25 \\
    Base w/ Context & 2.72 & 1.04 & 1.73 & 0.22 & 0.44 \\
    \midrule
    Ground Truth Encoded & 1.75 & 0.52 & 0.68 & 0.03 & 0.47 \\
    Ground Truth Raw & 0.00 & 0.00 & 0.00 & 0.00 & 0.80 \\
    \bottomrule
  \end{tabular}
  \caption{Synthesis model experiment results for \textbf{Cerberus4} Test with 620 examples.}
  \label{tab:results_cerberus4}
\end{table*}

\begin{table*}[h]
\sisetup{table-format=2.2,round-mode=places,round-precision=2,table-number-alignment = center,detect-weight=true,detect-inline-weight=math}
  \centering
  \begin{tabular}{l S S S S S}
    \toprule
    \multicolumn{1}{c}{\multirow{2}{*}{Model}} & \multicolumn{2}{c}{VGGish~\cite{hershey17audiocnn}} & \multicolumn{2}{c}{TRILL~\cite{trill}} & {\multirow{2}{*}{{Transcription F1 $\uparrow$}}} \\
    {} & {Recon. $\downarrow$} & {FAD $\downarrow$} & {Recon. $\downarrow$} & {FAD $\downarrow$} & {} \\  
    \midrule
    Small Autoregressive & 3.47 & 1.02 & 6.76 & 0.47 & 0.27 \\
    Small w/o Context & 3.62 & 1.01 & 7.57 & 0.46 & 0.22 \\
    Small w/ Context & 3.72 & 1.08 & 8.08 & 0.56 & 0.21 \\
    Base w/ Context & 2.93 & 0.97 & 2.39 & 0.35 & 0.30 \\
    \midrule
    Ground Truth Encoded & 1.97 & 0.43 & 1.04 & 0.03 & 0.27 \\
    Ground Truth Raw & 0.00 & 0.00 & 0.00 & 0.00 & 0.80 \\
    \bottomrule
  \end{tabular}
  \caption{Synthesis model experiment results for \textbf{MAESTROv3} Test with 177 examples.}
  \label{tab:results_maestrov3}
\end{table*}

\begin{table*}[h]
\sisetup{table-format=2.2,round-mode=places,round-precision=2,table-number-alignment = center,detect-weight=true,detect-inline-weight=math}
  \centering
  \begin{tabular}{l S S S S S}
    \toprule
    \multicolumn{1}{c}{\multirow{2}{*}{Model}} & \multicolumn{2}{c}{VGGish~\cite{hershey17audiocnn}} & \multicolumn{2}{c}{TRILL~\cite{trill}} & {\multirow{2}{*}{{Transcription F1 $\uparrow$}}} \\
    {} & {Recon. $\downarrow$} & {FAD $\downarrow$} & {Recon. $\downarrow$} & {FAD $\downarrow$} & {} \\  
    \midrule
    Small Autoregressive & 4.92 & 1.08 & 14.68 & 0.49 & 0.13 \\
    Small w/o Context & 4.11 & 1.08 & 8.21 & 0.54 & 0.24 \\
    Small w/ Context & 4.32 & 1.14 & 9.65 & 0.63 & 0.16 \\
    Base w/ Context & 3.74 & 1.04 & 3.29 & 0.33 & 0.25 \\
    \midrule
    Ground Truth Encoded & 2.24 & 0.46 & 0.98 & 0.03 & 0.21 \\
    Ground Truth Raw & 0.00 & 0.00 & 0.00 & 0.00 & 0.30 \\
    \bottomrule
  \end{tabular}
  \caption{Synthesis model experiment results for \textbf{MusicNet} Test with 15 examples.}
  \label{tab:results_musicnet}
\end{table*}

\begin{table*}[h]
\sisetup{table-format=2.2,round-mode=places,round-precision=2,table-number-alignment = center,detect-weight=true,detect-inline-weight=math}
  \centering
  \begin{tabular}{l S S S S S}
    \toprule
    \multicolumn{1}{c}{\multirow{2}{*}{Model}} & \multicolumn{2}{c}{VGGish~\cite{hershey17audiocnn}} & \multicolumn{2}{c}{TRILL~\cite{trill}} & {\multirow{2}{*}{{Transcription F1 $\uparrow$}}} \\
    {} & {Recon. $\downarrow$} & {FAD $\downarrow$} & {Recon. $\downarrow$} & {FAD $\downarrow$} & {} \\  
    \midrule
    Small Autoregressive & 3.43 & 1.10 & 4.99 & 0.46 & 0.30 \\
    Small w/o Context & 2.96 & 1.17 & 2.83 & 0.54 & 0.30 \\
    Small w/ Context & 3.02 & 1.17 & 2.87 & 0.54 & 0.24 \\
    Base w/ Context & 2.80 & 1.05 & 1.41 & 0.25 & 0.39 \\
    \midrule
    Ground Truth Encoded & 1.80 & 0.50 & 0.49 & 0.02 & 0.35 \\
    Ground Truth Raw & 0.00 & 0.00 & 0.00 & 0.00 & 0.59 \\
    \bottomrule
  \end{tabular}
  \caption{Synthesis model experiment results for \textbf{Slakh2100} Test with 146 examples.}
  \label{tab:results_slakh}
\end{table*}

\begin{table*}[h]
\sisetup{table-format=2.2,round-mode=places,round-precision=2,table-number-alignment = center,detect-weight=true,detect-inline-weight=math}
  \centering
  \begin{tabular}{l S S S S S}
    \toprule
    \multicolumn{1}{c}{\multirow{2}{*}{Model}} & \multicolumn{2}{c}{VGGish~\cite{hershey17audiocnn}} & \multicolumn{2}{c}{TRILL~\cite{trill}} & {\multirow{2}{*}{{Transcription F1 $\uparrow$}}} \\
    {} & {Recon. $\downarrow$} & {FAD $\downarrow$} & {Recon. $\downarrow$} & {FAD $\downarrow$} & {} \\  
    \midrule
    Small Autoregressive & 3.65 & 1.00 & 6.60 & 0.25 & 0.55 \\
    Small w/o Context & 3.68 & 1.06 & 7.72 & 0.35 & 0.49 \\
    Small w/ Context & 3.59 & 1.10 & 6.63 & 0.47 & 0.48 \\
    Base w/ Context & 3.35 & 1.03 & 4.09 & 0.26 & 0.52 \\
    \midrule
    Ground Truth Encoded & 2.60 & 0.61 & 3.33 & 0.09 & 0.59 \\
    Ground Truth Raw & 0.00 & 0.00 & 0.00 & 0.00 & 0.79 \\
    \bottomrule
  \end{tabular}
  \caption{Synthesis model experiment results for \textbf{Guitarset} Validation with 238 examples.}
  \label{tab:results_guitarset}
\end{table*}

\begin{table*}[h]
\sisetup{table-format=2.2,round-mode=places,round-precision=2,table-number-alignment = center,detect-weight=true,detect-inline-weight=math}
  \centering
  \begin{tabular}{l S S S S S}
    \toprule
    \multicolumn{1}{c}{\multirow{2}{*}{Model}} & \multicolumn{2}{c}{VGGish~\cite{hershey17audiocnn}} & \multicolumn{2}{c}{TRILL~\cite{trill}} & {\multirow{2}{*}{{Transcription F1 $\uparrow$}}} \\
    {} & {Recon. $\downarrow$} & {FAD $\downarrow$} & {Recon. $\downarrow$} & {FAD $\downarrow$} & {} \\  
    \midrule
    Small Autoregressive & 3.78 & 0.96 & 7.67 & 0.41 & 0.15 \\
    Small w/o Context & 3.67 & 0.98 & 7.56 & 0.47 & 0.23 \\
    Small w/ Context & 3.70 & 0.99 & 7.71 & 0.48 & 0.18 \\
    Base w/ Context & 3.21 & 0.86 & 3.54 & 0.23 & 0.27 \\
    \midrule
    Ground Truth Encoded & 2.50 & 0.54 & 2.69 & 0.07 & 0.28 \\
    Ground Truth Raw & 0.00 & 0.00 & 0.00 & 0.00 & 0.49 \\
    \bottomrule
  \end{tabular}
  \caption{Synthesis model experiment results for \textbf{URMP} Validation with 9 examples.}
  \label{tab:results_urmp}
\end{table*}

\end{document}